\documentclass[10pt,conference,compsocconf]{IEEEtran}
\usepackage{times}
\usepackage{enumerate}
\usepackage{amsmath}
\usepackage{amssymb}
\usepackage{graphicx}
\usepackage{caption}
\usepackage{cite}
\usepackage{url}

\usepackage[table]{xcolor}

\definecolor{tablegray}{rgb}{0.9, 0.9, 0.9}
\definecolor{decreasered}{rgb}{1, 0.7, 0.7}
\definecolor{increaseblue}{rgb}{0.8, 0.8, 1}
\definecolor{tableyellow}{rgb}{0.96, 0.96, 0}
\usepackage{array,booktabs,arydshln,xcolor}

\usepackage{lipsum}
\usepackage{dblfloatfix}

\ifCLASSINFOpdf
\else
\fi
\begin{document}
\pagenumbering{gobble}
%
\title{\textbf{\Large Backtracking (the) Algorithms on the Hamiltonian Cycle Problem}}



\author{\IEEEauthorblockN{~\\[-0.4ex]\large Joeri Sleegers\\[0.3ex]\normalsize}
\IEEEauthorblockA{Mice \& Man Software and A.I. Development \\
Amsterdam \\
The Netherlands\\
{\tt jsleegers@hotmail.com}}
\and
\IEEEauthorblockN{~\\[-0.4ex]\large Daan van den Berg\\ [0.3ex]\normalsize}
\IEEEauthorblockA{Yamasan Science \& Education \\
 Amsterdam\\ The Netherlands\\
 {\tt daan@yamasan.nl}}}


\maketitle

\begin{abstract}
Even though the Hamiltonian cycle problem is NP-complete, many of its problem instances are not. In fact, almost all the hard instances reside in one area: near the Koml\'os-Szemer\'edi bound, where randomly generated graphs have an approximate 50\% chance of being Hamiltonian. If the number of edges is either much higher or much lower, the problem is not hard -- most backtracking algorithms decide such instances in (near) polynomial time. Recently however, targeted search efforts have identified very hard Hamiltonian cycle problem instances \textit{very far away} from the Koml\'os-Szemer\'edi bound. In that study, the used backtracking algorithm was  Vandegriend-Culberson's, which was supposedly the most efficient of all Hamiltonian backtracking algorithms. In this paper, we make a unified large scale quantitative comparison for the best known backtracking algorithms described between 1877 and 2016. We confirm the suspicion that the Koml\'os-Szemer\'edi bound is a hard area for all backtracking algorithms, but also that Vandegriend-Culberson is indeed the most efficient algorithm, when expressed in consumed computing time. When measured in recursive effectiveness however, the algorithm by Frank Rubin, almost half a century old, performs best. In a more general algorithmic assessment, we conjecture that edge pruning and non-Hamiltonicity checks might be largely responsible for these recursive savings.  When expressed in system time however, denser problem instances require much more time per recursion. This is most likely due to the costliness of the extra search pruning procedures, which are relatively elaborate. We supply large amounts of experimental data, and a unified single-program implementation for all six algorithms. All data and algorithmic source code is made public for further use by our colleagues.

\end{abstract}


\begin{IEEEkeywords}
\textit{Hamiltonian Cycle; exact algorithm; exhaustive algorithm; heuristic; phase transition; order parameter; data analytics; instance hardness; replication}.%
\end{IEEEkeywords}

%
\IEEEpeerreviewmaketitle

\section{Preamble}
\noindent Traversing a crack in the fabric of the scientific spacetime-continuum, this paper finds itself in the unusual position that its designated  conclusions have already been overthrown. Following a replication study \cite{van2018predictive}, these extended results should  have been published earlier, but as history unfolded, they simply were not. In any case, the study by Cheeseman, Kanefsky \& Taylor (henceforth: `Cetal') was first \cite{cheeseman1991really}. Sharpening the resolution of the $P\stackrel{?}{=}NP$ problem, they showed that for various NP-complete problems, Hamiltonian cycle, graph colouring, satisfiability and the asymmetric traveling salesman problem\footnote{Even though the authors themselves dubbed traveling salesman as ``NP-complete'', they solved the NP-hard version of the problem.}, instances vary greatly in their computational hardness. Sporting over 1400 citations, the paper became a landmark in the field.

\begin{figure*}[ht]
\centering%
\includegraphics[width=\linewidth]{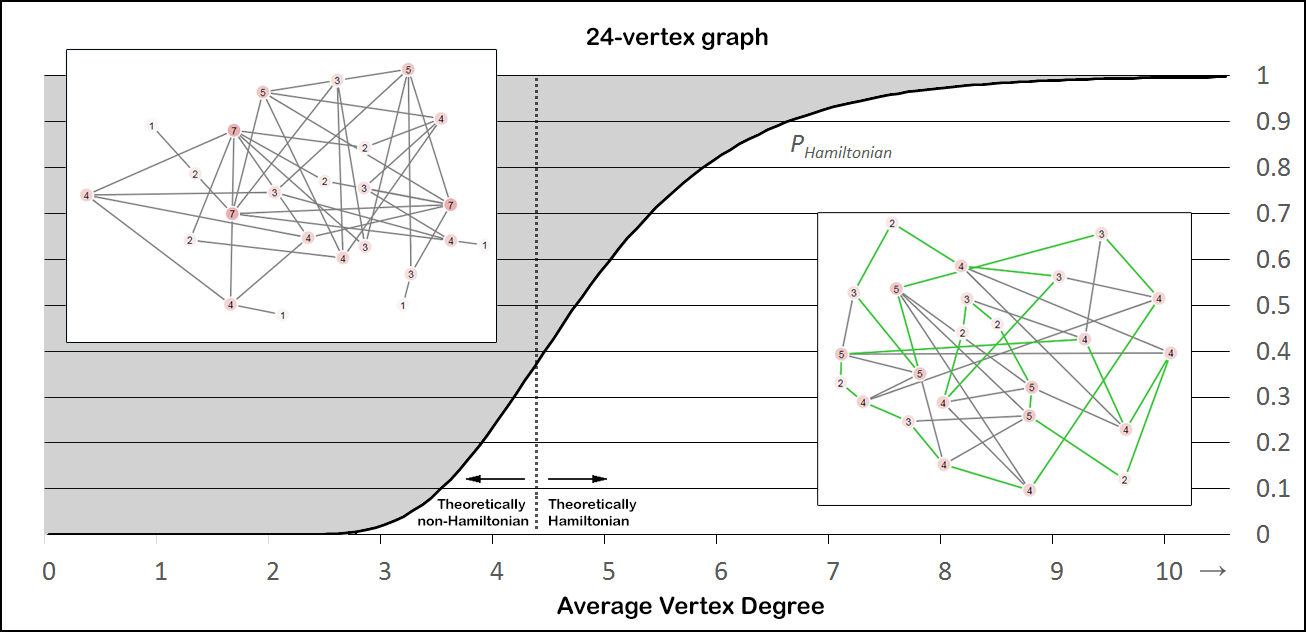}
\caption{The probability of a randomly generated graph being Hamiltonian depends on the average vertex degree, and is  sigmoidally shaped around the ‘threshold point’ of $ln(V) + ln(ln(V))$. Top-left inset is a non-Hamiltonian random graph, bottom-right inset is a Hamiltonian graph with the Hamiltonian cycle itself being highlighted.}
\label{fig:hamiltonianphasetransition}
\end{figure*}

It took nearly 30 years for Cetal's results on traveling salesman to be overthrown, appearing to have been flawed by an overlooked roundoff error\cite{sleegers2020really} (also see the accompanying videos: \cite{videoATSP1}\cite{videoATSP2}). Their results on the Hamiltonian cycle problem however, were succesfully replicated and published at IARIA's Data Analytics 2018 conference, and brought the data, sourcecode and figures alive in online interactively publicly accessible resources \cite{van2018predictive}\cite{Gijsinteractivesite}. The results of that extended study showed that for Cetal's algorithm, and two others, the hardest instances of the Hamiltonian cycle were located along an area known as the  ``Koml\'os-Szemer\'edi bound''. So far so good.

But the lingering question was how large the influence of the solvers was. Were the found hard instances hard specifically for the solving algorithms used? For the pruning methods? For the branching heuristic? And to make matters worse, a recent follow-up study showed that for the allegedly most efficient Hamiltonian cycle backtracker, the Vandegriend-Culberson (henceforth: `Vacul')-algorithm, (which still requires exponential time in the worst case), the hardest instances were located \textit{very  far away} from the Koml\'os-Szemer\'edi bound \cite{sleegers2020plant}\cite{Sleegers2020Looking}. Only findable by sophisticated evolutionary algorithms and a whole lot of computing power, these very hard instances never showed up in earlier studies.

So is there an insurmountable contradiction here? Probably not. The authors conjecture three reasons for their counterintuitive results: ``A first explanation for these surprising results is that these results are specific for the backtracking algorithm we used. This is unlikely however, as the algorithm minimizes most other backtracking algorithms found in literature (yet unpublished results)''. Put differently: chances are very high that these graphs are \textit{also} hard for other complete backtracking algorithms but evidence pending, it remained unconfirmed as yet. This study will at the very least post a very serious sidenote to that hypothesis. A second explanation given by these authors might be that in most studies on Hamiltonian cycle backtracking algorithms, runs are cutoff after a preset number of recursions, as even small graphs can take up significant decision time. Although theoretically feasible, in practice these cutoff points are usually situated near the Koml´os-Szemer´edi bound, and not in edge-dense regions far away where the extremely hard instances were located.

The third and most compelling explanation might therefore come from their obsevation that on first glance, these very hard instances might have low Kolmogorov complexity – they are in some sense \textit{structured} graphs. And as unstructured objects in randomly generated ensemble vastly outnumber the structured objects, the chances of being created by a stochastic process (which is the case in most large-scale comparative studies) are extremely small. Put differently: one would simply not find the very hard instances unless knowing exactly where to look. Or as the authors more poetically phrased: ``These graphs are an isolated island of structured hardness in a ocean of unstructured easiness. Whether more such islands exist, and what they look like, awaits further exploration.''\footnote{The literate reader might remind Aldous Huxley's famous quote \textit{``Consider the horse. They considered it.'' -- A Brave New World, 1932.} \cite{huxley2007brave}}   \cite{sleegers2020plant}.

So to understand their results, a couple of things need to happen. First, Vacul's algorithm must \textit{evidently} minimize the other backtracking algorithms. Then, a global evolutionary search algorithm should look for the hardest instances for \textit{all} these backtracking algorithms. Third, a hardness hierarchy should be made for all six algorithms, another large quantitative study. Fourth, an aggregated theory should explain \textit{why} some instances of the Hamiltonian cycle problem are harder than others, and what their relation to the Koml\'os-Szemer\'edi bound is (or is not!). It is the authors' belief that such an aggregated theory exists. Beyond that, as a fifth point, might lie implications for computability, and the $P\stackrel{?}{=}NP$ problem itself, but developments in this area will depend on foregoing results, and are as yet too close to call.

In this paper, which is an extension of the IARIA'18 paper mentioned in the second paragraph \cite{van2018predictive}, we will address step one: show that in large random ensembles, the hardest instances of the Hamiltonian cycle problem lie around the Koml\'os-Szemer\'edi bound for \textit{all} well-known general backtracking algorithms found in literature. We will make exact calculations on large ensembles, and perform a rigorous comparative analysis. While reading the paper, the reader should keep two things in mind: first, other backtracking algorithms than those in this study are well existable and second: it is possible that much harder instances exist for any backtracker in this study or elsewhere. As shown earlier, extremely hard instances are likely extremely rare, extremely hard to find, but also extremely important for the $P\stackrel{?}{=}NP$ problem, as it is these instances that etch upper bounds on  these algorithms' runtimes. Finding these, possibly with targeted evolutionary algorithms, is the second step and will hopefully be done in the near future. For now, we will work with completely unbiased random ensembles of instances to map out the gross features of the complexity landscape.

\begin{figure}[h]
\centering%
\includegraphics[width=\columnwidth]{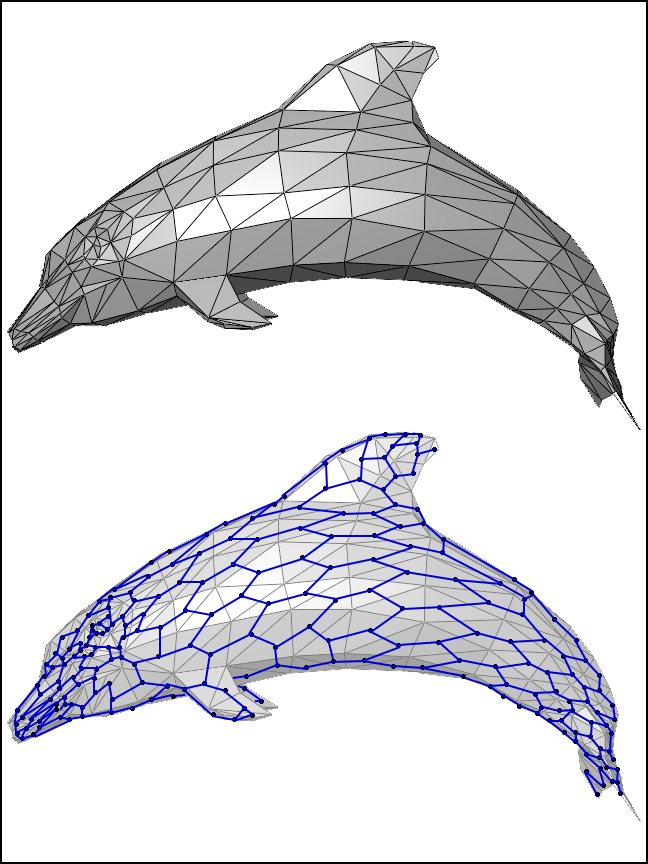}
\caption{Fast rendering of triangle mesh 3D images critically depends on finding Hamiltonian cycles through the corresponding 'cubic' graphs, in which every vertex has a maximum degree of three.}\label{fig:trianglemesh}
\end{figure}

\section{Introduction}
\noindent The ''Great Divide'' between P and NP has haunted computer science and related disciplines for over half a century. Problems in P are problems for which the runtime of the best known algorithm increases polynomially with the problem size, for example, calculating the average of an array of numbers. If the array doubles in size, so does the runtime of the best known algorithm - a polynomial increase. A problem in NP however, has no such polynomial-time algorithm and it is an open question whether one will ever be found. An example hereof is 'satisfiability' (sometimes abbreviated to SAT), in which an algorithm assigns values 'true' or 'false' to variables in Boolean formulas like $( a \lor \lnot b \lor d ) \land ( b \lor c \lor \lnot d )$. The task is to choose the variable assignment so that the formula as a whole is satisfied (becomes 'true'), \textit{and returning that assignment}, or making sure that no such assignment exists. Algorithms that do this, algorithms that are guaranteed to give a solution whenever it exists and return ‘no’ otherwise, are called \textit{exact} algorithms\footnote{Technically speaking, exact algorithms for decision problems such as the Hamiltonian cycle problem should be called \textit{complete}. But since `exact' seems more in schwung recently, we will stick to that.}

Being exact is a great virtue for an algorithm, but it comes at a hefty price. Often, these algorithms operate by \textit{brute-force} procedures: exhaustively trying all combinations for all variables until a solution is found, which usually takes vast amounts of time. Depth-first search is the keystone example of this study; it is exhaustively exact and indeed consumes harrowing amounts of computational power (see Figures \ref{fig:zesluik_recursions} and \ref{fig:zesluik_time}). Smarter algorithms exist too; clever search pruning can speed things up by excluding large sections of state space at the cost of some extra computational instructions, an investment that usually pays off. Heuristic algorithms are also fast but not necessarily exact - so it is not guaranteed a solution is found if one exists. After decades of research, the runtimes of even the most efficient complete SAT-algorithm known today still increases exponentially with the number of variables -- much worse than polynomial, even for low exponents. Therefore, SAT is in NP, a class of `Notorious Problems' that rapidly become unsolvable as their size increases. In practice, this means that satisfiability problems (and other problems in NP) with only a few hundred variables are practically unsolvable, whereas industries such as chip manufacture or program verification in software engineering could typically employ millions ˜\cite{bryant1991complexity}˜\cite{ivanvcic2008efficient}.  

The problem class NP might be considered ``the class of dashed hopes and idle dreams'' \cite{dashed}, but nonetheless scientists managed to pry loose a few bricks in the great wall that separates P from NP. Most notably, the seminal work ''Where the \textit{Really} Hard Problems Are'' by Cheeseman, Kanefsky and Taylor (henceforth abbreviated to 'Cetal'), showed that although runtime increases non-polynomially for problems in NP, some \textit{instances} of these hard problems might actually be easy to solve ˜\cite{cheeseman1991really}. Not every formula in SAT is hard -- easily satisfiable formulas exist too, even with many variables, but the hard ones keep the problem as a whole in NP. But Cetal's great contribution was not only to expose the huge differences in instance hardness within a single NP-problem, they also showed \textit{where} those really hard instances are -- and how to get there. Their findings were followed up numerous times and truly exposed some of the intricate inner anatomy of instance hardness, and problem class hardness as a whole.

\begin{figure*}[h]
\centering%
\includegraphics[width=\linewidth]{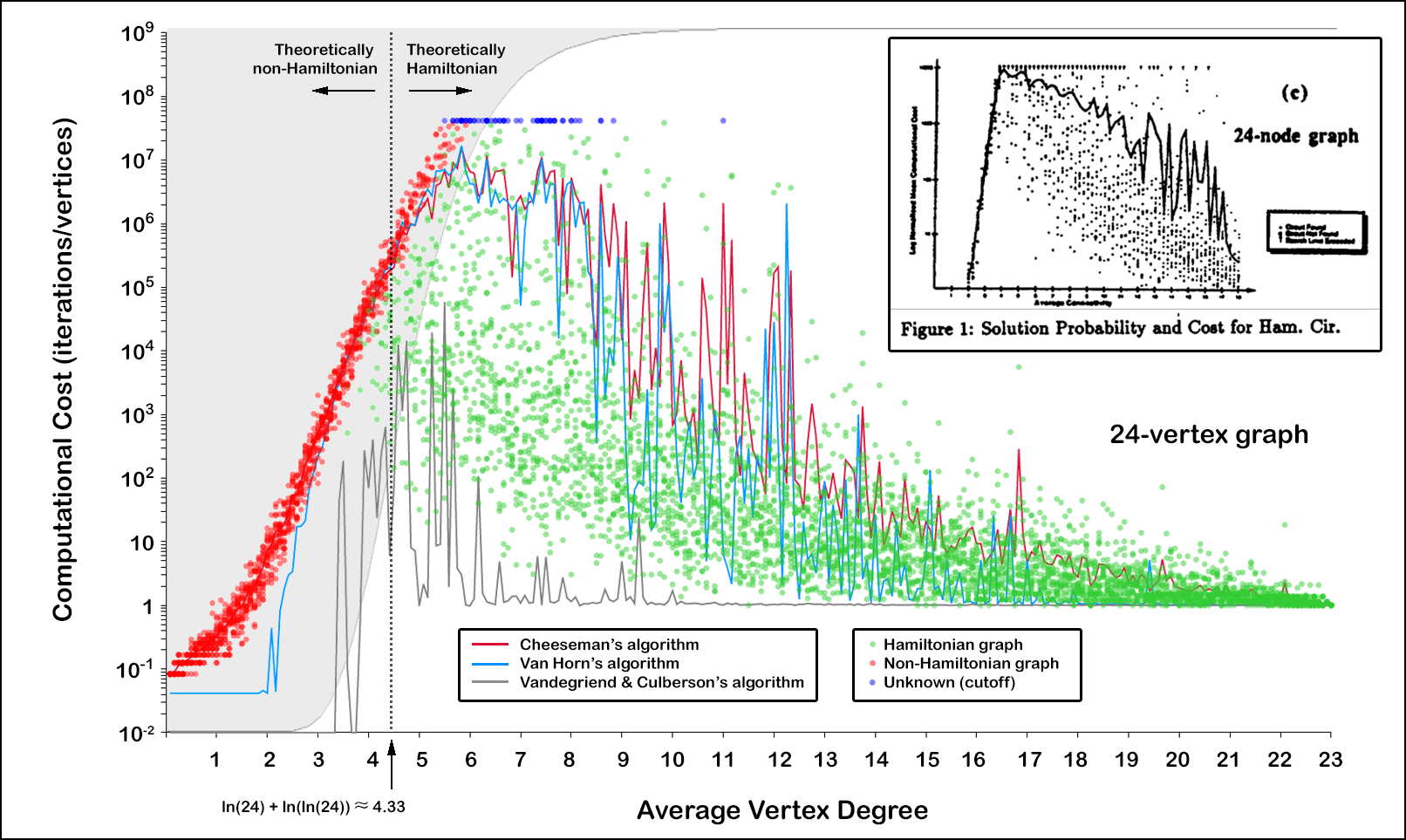} 
\caption{Results from an earlier replication of Cetal's seminal work on Hamiltonian cycle hardness, extended with algorithms by Van Horn and Vacul. The top-right inset is Cetal’s original figure, and it covers no data points. Note how the computational cost is highest along the Koml\'os-Szemer\'edi bound.}
\label{fig:vanhornsresults}
\end{figure*}

So where \textit{are} these hard problem instances then? According to Cetal, they are hiding in the phase transition. For the problems in their study, instances suddenly jump from `having many solutions' to `having no solutions' when their constrainedness changes. For an example in satisfiability, most randomly generated SAT-formulas of two clauses and four variables such as our formula $( a \lor \lnot b \lor d ) \land ( b \lor c \lor \lnot d )$  are easily satisfiable; they have many assignments that make them true. But as soon as the order parameter, the ratio of clauses versus variables $\alpha$, passes $\approx 4.26$, (almost) no satisfiable formulas exist ~\cite{larrabee1992evidence}˜\cite{kirkpatrick1994critical}. So if we randomly generate a formula with 20 or more clauses on these same four variables, it is almost certainly unsatisfiable and those rare formulas that \textit{are} satisfiable beyond the phase transition have very few solutions -- which counterintuitively enough makes them easy again. So, for most exact algorithms, both extremes are quickly decided: for the highly satisfiable formulas in $\alpha<< 4.26$, a solution is quickly found, and unsatisfiable formulas in $\alpha>> 4.26$ are quickly proven as such. But in between, just around $\alpha = 4.26$, where the transition from satisfiable to unsatisfiable takes place, are formulas that take the longest to decide upon. This is where the really hard problem instances are: hiding in the phase transition that separates the solvable from the unsolvable region. 

Cetal identify this order parameter not only for SAT; the Hamiltonian cycle problem has one too, and so does k-colorability. For the Hamiltonian cycle problem though, one should understand that the phase transition is somewhat inverse to SAT: as one adds edges to a graph, it becomes \textit{solvable} rather than unsolvable. But the general principle still holds: both extremes are easy, and the phase transition is where the really hard problem instances are. And although Cetal's seminal results are relatively coarse, they have been followed up in more detail, and they are solid ˜\cite{kirkpatrick1994critical,gent1994easy,hogg1994hardest,hogg1996refining}. To put it in a quote by Ian Gent and Toby Walsh: ''[Indeed, we have yet to find an NP-complete problem that \textit{lacks} a phase transition]'' ˜\cite{gent1996tsp}.

The ubiquity of phase transitions throughout the class is not a complete surprise. Satisfiability, k-colorability and the Hamiltonian cycle problem are \textit{NP-complete} problems; a subset of problems in NP that with more or less effort can be transformed into each other ˜\cite{garey2002computers}. This means a lot. This means that if someone finds a polynomial exact algorithm for just one of these problems, all of them become easy and the whole hardness class will simply evaporate. That person would also be an instant millionaire thanks to the Clay Mathematics Institute that listed the P$\stackrel{?}{=}$NP-question as one of their `Millenium Problems' ˜\cite{jaffe2006millennium}. But the intricate relations inside NP-completeness might also stretch into the properties of phase transitions and instance hardness. Or, to pour it into another fluid expression by Ian Gent and Toby Walsh ``[Although any NP-complete problem can be transformed into any other NP-complete problem, this mapping does \textit{not} map the problem space uniformly]'' ˜\cite{gent1996tsp}. So, a phase transition in say, satisfiability, does \textit{not} guarantee the existence of a phase transition in Hamiltonian Cycle or in Vertex Coloring. The fact is though, that Cetal do find them for all three, and their results are solid.

\section{The Hamiltonian Phase Transition}

\noindent The Hamiltonian cycle problem comes in many different shapes and forms, but in its most elementary formulation involves finding a path (a sequence of distinct edges) in an undirected and unweighted graph that visits every vertex exactly once, and forms a closed loop. The probability of a random graph being Hamiltonian (i.e., having a Hamiltonian Cycle), has been thoroughly studied ˜\cite{erdos1960evolution,posa1976hamiltonian,komlos1983limit}. In the limit, it is a smooth function of vertex degree and therefore the probability for a random graph of $v$ Vertices and $e$ edges being Hamiltonian can be calculated analytically\footnote{An unfortunate convention: please note that the $e$ in the right hand side of Eq. \ref{eq1} is the base of the natural logarithm, whereas in the left hand side, $e$ is the number of edges}:
\begin{equation}
		P_{Hamiltonian}(v,e) = e^{-e^{-2c}}
\label{eq1}
\end{equation}
in which
\begin{equation}
		e = \frac{1}{2} v\cdot ln(v) + \frac{1}{2} v\cdot ln( ln(v)) + c \cdot v
\label{eq2}
\end{equation}
Like the phase transition around $\alpha$ in SAT, the Hamiltonian phase transition is also sigmoidally shaped\footnote{Remember this transition is \textit{invertalized} with respect to SAT: it goes `from no to yes' when getting denser, where SAT goes `from yes to no'.} across a ‘threshold point’, the average degree of $ln(v) + ln(ln(v))$  for a graph of $v$ vertices (Figure \ref{fig:hamiltonianphasetransition}). The phase transition gets ever steeper for larger graphs and becomes instantaneous at the threshold point as $v$ goes to infinity. For this (theoretical) reason, the probability of being Hamiltonian at the threshold point is somewhat below 0.5 at $e^{-1} \approx 0.368$.

The probability of being Hamiltonian is one thing, deciding whether a given graph actually \textit{has} a Hamiltonian cycle is quite another. A great number of exact algorithms have been developed through the years, the earliest being exhaustive methods that could run in $O(v!)$ time~\cite{roberts1966systematic}. A dynamic programming approach, quite advanced for the time, running in $O(n^2 2^n)$ was built by by Michael Held \& Richard Karp, and by Richard Bellman independently~\cite{held1962dynamic}~\cite{bellman1962dynamic}. Some early edge pruning efforts and check routines can be found in the work of Silvano Martello and Frank Rubin whose algorithms still run in $O(v!)$ but are in practice much faster (as we will show in Figures \ref{fig:zesluik_recursions} and \ref{fig:zesluik_time}) ~\cite{martello1983algorithm}~\cite{rubin1974search}. Many of their techniques eventually ended up in the algorithm by Vandegriend \& Culberson  ~\cite{vandegriend1998gn}. Algorithms by Bollob\'as and Bj\"orklund run faster than Bellman--Held--Karp, but are technically speaking not exact for finite graphs~\cite{bollobas1987algorithm}~\cite{bjorklund2014determinant}. The 2007 algorithm by Iwama \& Nakashima ~\cite{iwama2007improved} runs in $O(2^{1.251v})$ time on cubic graphs, thereby improving Eppstein's 2003 algorithm that runs in $O(2^{1.260v})$. While these kind of marginal improvements on specific instances are typical for the progress in the field, these two actually deserve some extra attention. 

The cubic graph, in which every vertex has a maximum degree of three, is of special importance in the generation of 3D computer images. Many such images are built up from triangle meshes, and as specialized hardware render and shade triangles at low latencies, the performance bottleneck is actually in feeding the triangular structure into the hardware. A significant speedup can be achieved by not feeding every triangle by itself, but by combining them into triangle strips. An adjacent triangle can be defined by only one new point from the previously fed triangle, and therefore adjacent triangles combined in a single strip can speedup the feeding procedure by a maximum factor three for each 3D object. Finding a single strip that incorporates all triangles in the mesh is equivalent to finding a Hamiltonian cycle through the corresponding cubic graph in which every triangle is a vertex, which makes both Eppstein's and Iwama\&Nakashima's result of crucial importance for the 3D imagery business (see Figure~\ref{fig:trianglemesh}).

So concludingly, none of the exact algorithms for the Hamiltonian cycle problem runs faster than exponential on all instances (with vertices of any degree), and the Bellman--Held--Karp ``is still the strongest known'', as Andreas Bj\"orklund states on the first page of his 2010-paper~\cite{bjorklund2014determinant}. Summarizingly, all backtracking algorithms still run in $O(v!)$ time. In the next sections, we will have a closer look at the algorithmic details of all \textit{exact} algorithms from previous pargraphs. We will first make a structural and historical comparison, and after that bang out a lot of data to make a rigorous quantitative comparison as well. Finally, we will draw some conclusions, discuss the impact of the work, and project a trajectory for future research.

\section{Algorithmics}
\noindent For this extended investigation, we generated large numbers of problem random instances for the Hamiltonian cycle problem, varying in numbers of vertices and edges. We then solved these using almost every authoritative exact algorithm we could find: \textbf{Depth-first search} invented far before any modern day computer in 1882, \textbf{Rubin's algorithm} from 1974, \textbf{Martello's algorithm '595'} published in 1983, \textbf{Cetal's} algorithm,  used in their 1991 seminal work on instance hardness, Vandegriend \& Culberson's, abbreviated to \textbf{Vacul's algorithm}, an elaborate machinery published in 1998 and finally \textbf{Van Horn's algorithm}, directly derived from Cetal's, published in 2018 \cite{even1979graph}\cite{rubin1974search}\cite{martello1983algorithm}\cite{cheeseman1991really}\cite{vandegriend1998gn}\cite{van2018predictive}.

There are at least several more, some of which are very old and without Google Scholar index or available pdfs. One prominent candidate missing in our list is the dynamic programming implementation built by Michael Held \& Richard Karp, and by Richard Bellman independently \cite{held1962dynamic}\cite{bellman1962dynamic}. While the algorithm is complete, and with a time complexity of $O(v^2 2^v)$ has a better worst case performance than our six algorithms which all have $O(v!)$, it has significant memory requirements. Furthermore, it is not a backtracker and therefore does not perform `recursions' as such, making a direct comparison to the other six slightly more difficult. It might be a viable candidate for a future comparison but for now, we will stick to backtracking algorithms, and canalize towards a generalized approach.

As it turns out, these six algorithms have several similarities and differences. But even though the respective papers have significant numbers of citations, some of the (especially earlier) authors of these particular algorithms seemed to be unaware of each others' progress. The main exception is Vacul, who include a reference to both Martello and Cetal, but missed Frank Rubin's work, which might simply be due to geographical dispersity and the lack of internet. Or conversely: the relatively recent development of computational resources, and  proliferation of global communication enables us \textit{now} to make a large structural comparison between them relatively easily. In any case, there appear to be some globally emerging algorithmic design patterns for this problem, which are dispersedly found accross algorithms. We will structurally compare all, and assess their effectiveness regarding the Hamiltonian cycle problem. 

In this study we generalize the approach, unifying similar procedural subroutines to the same piece of source code. The only feature that was removed is the \text{random restart} option from Vacul's algorithm. Surely a good way of shortening the average runtime on many NP-complete problems \cite{gomes1999fine}, it also makes one-off comparisons a lot harder on large randomized instance ensembles such as ours. As most algorithms predate the internet, none came with readily implementable source code. We like to emphasize our belief that text such as you are reading now is an inferior medium for communicating algorithmics. As the smallest of details can make the largest differences when it comes to runtimes. Therefore, one should \textit{always supply publicly accessible source code} when writing about algorithms. We will try to set the example by supplying ours \cite{sourcecodejoeri}. It is quite possible that algorithmic details as we chose them are different from the original authors' implementations. It is also thinkable that even where we are precise,  further improvements are possible. In any case, let us go forward with public source code along our publications on algorithms. It is better than text.

In the next section, we will describe the six backtrack algorithms used, but also the subroutines, many of which are common to multiple algorithms. The subjects are somewhat `interwoven' between the algorithmic explanation, simply because it seems to make the most sense storywise. Operationally speaking, it makes more sense to compartmentalize these subroutines, so that is how the reader will find them in the source code \cite{sourcecodejoeri}. For a general overview, please refer to Table \ref{table:techniques}.

\begin{table*}[hb]
\small
     \centering
 \begin{tabular}{|c|c|c|c|}
 \hline
     \rowcolor{tableyellow}\textbf{Algorithm} & \textbf{Pruning} & \textbf{Heuristic} & \textbf{Non-Hamiltonicity} \textbf{Checks}\\
     \hline
     \hline
      \rowcolor{tablegray}Depth-First & None & None & None \\
      \rowcolor{tablegray}Cetal's & None & High & None        \\
      \rowcolor{tablegray}Van Horn's & None & Low & None            \\
      \rowcolor{tablegray}Martello's & Solution, Path & Low & Degree          \\
      \rowcolor{tablegray}Rubin's & Solution, Path, Neighbour & None   & Degree, One-Connectedness, Disconnectedness, Premature Closure          \\
      \rowcolor{tablegray}Vacul's & Solution, Path, Neighbour & Low & Degree, One-Connectedness, Disconnectedness \\
 \hline
 \end{tabular}
 
 \caption{Overview of all the techniques used by the algorithms that are examined in this research. For the pruning methods, it is displayed if an algorithm uses that method. For the branching heuristic it is presented if an algorithm uses one and if so which one.}
 \label{table:techniques}
\end{table*}

\section{Three Basic Algorithms}

\noindent \textbf{Depth-first search} can in every way be considered as the basis for all algorithms in this study. It is surely the oldest, gaining widespread popularity from Tarjan's paper \cite{tarjan1972depth}, but was designed roughly a century earlier by Frenchman Charles Pierre Tr\'{e}maux, and mentioned in a publication from 1882 as a strategy for solving mazes \cite{lucas1882recreations}. As (planar) mazes are in many ways equivalent to (planar) graphs, its success as an algorithm for graph traversal is hardly surprising.

In modern-day computers, depth-first search is readily implemented either by a recursive function or via a stack data structure, the latter of which is usually a constant factor slower. Depth-first search is an exact algorithm in optimization problems, meaning it will always return the best possible answer (e.g. the shortest route, or the optimal timetable). For decision problems like the Hamiltonian cycle problem or satisfiability, it will always return a solution if the problem instance has one, or ensure it does not exist. For finding a Hamiltonian cycle (or any other particular vertex order) in a graph, its ominous runtime complexity is $O(v!)$ in the number of vertices $v$, which makes it practically unusable for any number of $v$ over two digits. The algorithm can just as well be deployed for solving SAT-formulas, running in slightly less daunting exponential complexity of $O(2^n)$ in $n$, the number of Boolean variables.

\begin{figure*}[h]
    \centering
    \includegraphics[scale=0.68]{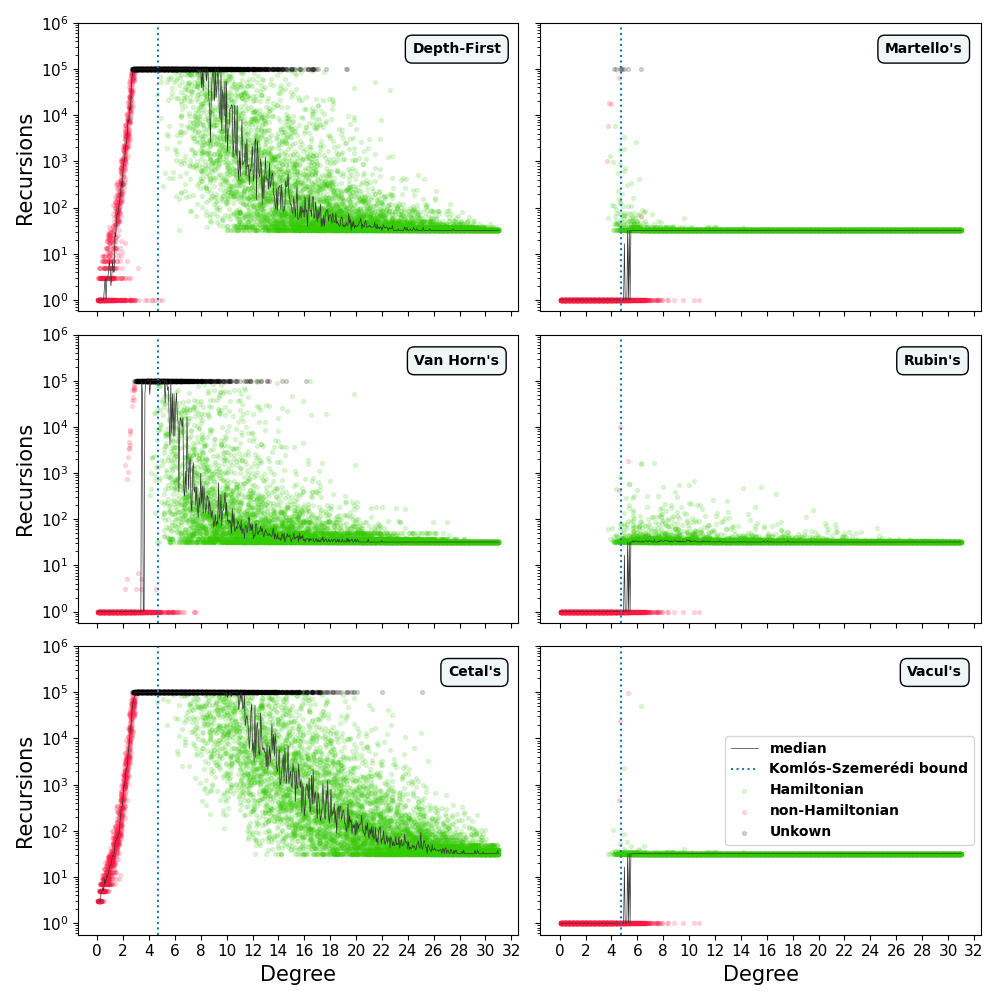}
    \caption{The number of recursions required for solving the 9920 random graphs of $v=32$. For all algorithms, the hardest graphs are situated close to the Koml\'os-Szemer\'edi bound of $ln(32)+ln(ln(32)) \approx 4.71$ where the probability of being Hamiltonian transitions from zero to one. But while the choice of branching heuristic clearly makes a difference (Depth-first, Van Horn's, Cetal's) the improvement is most dramatic when the algorithm also implements edge pruning and non-Hamiltonicity check procedures (Martello's, Vacul's, Rubin's). }
    \label{fig:zesluik_recursions}
\end{figure*}

\begin{figure*}[h]
    \centering
    \includegraphics[scale=0.68]{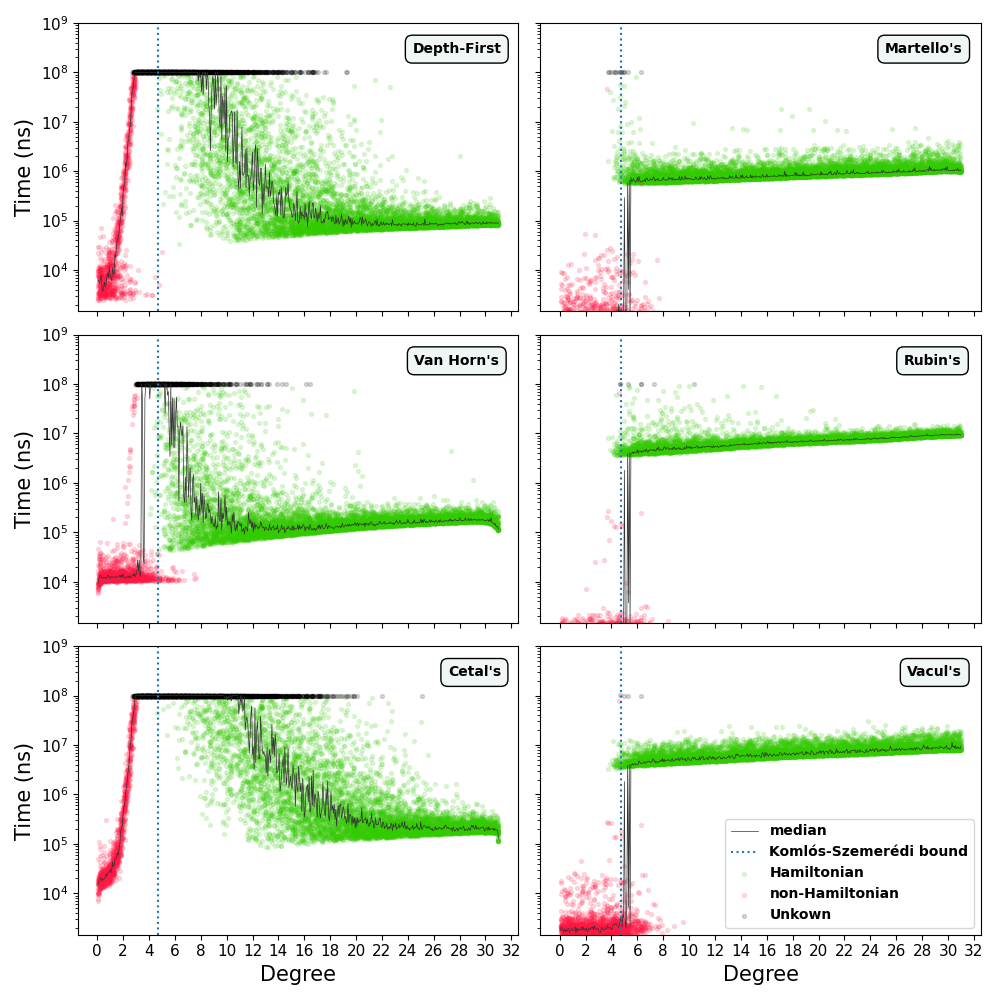}
    \caption{Procedures for pruning and non-Hamiltonicity checks (Martello's, Vacul's, Rubin's) clearly take `real' time, but even on graphs as small as these, they pay off -- the right hand side triplets have close to zero cutoffs. In these algorithms, the dramatic difference in time consumption between the red and green dots around the Koml\'os-Szemer\'edi bound show that these procedures save time especially on \textit{non-Hamiltonian} graphs.}
    \label{fig:zesluik_time}
\end{figure*}

Depth-first search is a constructive algorithm that starts at the first vertex in the data structure that holds the randomly generated graph, thereby not having any specified preference for degree. From there, its recursive step is to add the first adjacent vertex (again, in the order of the data structure) that is not already in the path. This step is repeated either until no adjacent vertices are available. This can mean two things: either all vertices are in the path, and a final check for closure confirms the existence of a Hamiltonian Cycle after which the algorithm halts, or the algorithm backtracks, removing the last vertex from the path and adding the next adjacent vertex in its place. If none such vertex exists it backtracks again. If this happens at the root level then all possibilities are exhausted, and the algorithm halts and returns "no Hamiltonian Cycle". As a typical property of (uninformed) exhaustive exact algorithms, it thereby tries all permutations of vertices if necessary and has a time complexity, or worst-case runtime, of $O(v!)$. 

A big \textit{practical} difference however, materializes by not just adding the next vertex from the data structure, but preferring either vertices of high degree or low degree. This can be done runtime, or by sorting the data structure upon reading the graph ahead of recusing. In our implementation, we always chose the former option, resorting during the run. As this next-vertex-preference is a somewhat rule-of-the-thumb, we will call it the \textbf{branching heuristic} and can be instantiated with categorical\footnote{Sometimes referred to as `symbolic parameter' or `qualitative parameter'.} parameter values$\{none, high, low\}$ (see Table \ref{table:techniques}). In its simplest implementation, it involves only changing a `$>$' to a `$<$' in the source code of the algorithm. Nonetheless the impact on the algorithm's performance for large ensembles of instances such as ours can be enormous \cite{van2018predictive}.

\textbf{Cetal's algorithm} follows the exact same paradigmic lines as plain depth-first search, but prefers higher degree vertices when branching. Ties are unaddressed, and whether a more sophisticated order of preference has any significant impact on the algorithm's runtime remains an open question, especially for larger graphs. Cetal's algorithm still runs in $O(v!)$, but its somewhat more fine-grained time complexity might be $O(v \cdot log(v))!$, with the additional term accounting for sorting the vertices to descending degree numbers. Cetal's paper was published with four experiments, among which the Hamiltonian cycle problem, and like this study, contained experimental results on an ensemble of random graphs in regular degree intervals.

\textbf{Van Horn's algorithm} is in many ways the opposite of Cetal's algorithm, starting at the vertex with the lowest degree and preferring lower degree vertices over higher degree vertices when recursing. It therefore also runs in $O(v!)$ time, but its  fine-grained time complexity is $v \cdot log(v)+v!$. Its only algorithmic difference is that it traverses the data structure backwardly, or alternatively, that its data structure is sorted in inverse order. Van Horn's algorithm was introduced during an extended replication of Cetal's Hamiltonian study, and thereby contained experimental results on random graphs of various average degree. The ensemble was likely larger than Cetal's\footnote{Cetal's ensemble size is unknown, but an estimate is given by Van Horn et al.}, and made publicly available, along with the source code of the algorithms used\footnote{We provide no reference to their source code to avoid confusion. Our implementation is in many ways a generalization of theirs. Of course, they do own the credit for open sourcing their implementation}.

\begin{table*}[hb]
    
     \centering
 \begin{tabular}{|c|c|c|c|c|c|}
 \hline
     \rowcolor{decreasered}\textbf{Algorithm} & \textbf{\#cutoff} & \textbf{\#finished} & \textbf{avg. recs} & \textbf{stddev. recs} & \textbf{total (x1000)}\\
     \hline
     \hline
      \rowcolor{tablegray}Depth-First & 1959 & 7961 & 2992 & 11661 & 219,716 \\
      \rowcolor{tablegray}Van Horn's & 814 & 9106 & 844 & 5890 & 8,908\\
      \rowcolor{tablegray}Cetal's & 2571& 7349 & 4551 & 13980 & 29,054 \\
      \rowcolor{tablegray}Martello's & 8 & 9912 & 40 & 686 & 1,193 \\
      \rowcolor{tablegray}Rubin's & 0 & 9920 & 31 & 106 & 307 \\
      \rowcolor{tablegray}Vacul's & 0 & 9920 & 44 & 1136 & 438  \\
 \hline
 \end{tabular}
 
 \caption{Recursions required by all six algorithms on the entire ensemble of random graphs. Note that `average recursions' and 'stddev recursions' apply to the \textit{finished} problem instances only.}
 \label{table:results_recursions}
\end{table*}

\begin{table*}[hb]
\small
     \centering
 \begin{tabular}{|c|c|c|c|c|c|}
 \hline
     \rowcolor{increaseblue}\textbf{Algorithm} & \textbf{\#cutoff} & \textbf{\#finished} & \textbf{avg. ns} & \textbf{stddev. ns} & \textbf{total (x 1M)}\\
     \hline
     \hline
      \rowcolor{tablegray}Depth-First & 1916 & 8004 & 3054626 & 11341555 & 216,049 \\
      \rowcolor{tablegray}Van Horn's & 813 & 9107 & 913606 & 5524056 & 89,620\\
      \rowcolor{tablegray}Cetal's & 2583& 7337 &  4695537 & 13464862 & 292,751 \\
      \rowcolor{tablegray}Martello's & 14 & 9906 & 864532 & 1502152 & 9,964 \\
      \rowcolor{tablegray}Rubin's & 7 & 9913 & 6080907 & 4439895 & 60,980 \\
      \rowcolor{tablegray}Vacul's & 4 & 9916 & 6125890 &  3718473 & 61,144  \\
 \hline
 \end{tabular}
 
 \caption{Runtime required by all six algorithms on the entire ensemble of random graphs. Average and stddev apply to the \textit{finished} problem instances only.}
 \label{table:results_time}
\end{table*}

\section{Search Pruning}

\subsection{Search Pruning: Edge Pruning}

\noindent One of the most prolific enhancements of depth-first search on NP-complete problems is \textbf{search pruning}: cutting off branches from the search tree that cannot hold a solution. If a contradictive pair of clauses is found from some variable assignment in CNF3SAT, the search backtracks immediately, departing from further assignments in the subtree. If a perfect rectangle packing problem contains an unfillable gap, it cannot be solved regardless of the rest of the configuration, so depth-first will halt and backtrack \cite{van2016almost}. As such, search pruning can be a valuable way of saving recursions, but the (sometimes huge) drawback is that the pruning procedure \textit{itself} takes computational resources too. Even though most pruning procedures are subexponential routines, they can theoretically be invoked in every recursion. The art therefore, is to balance an investment in pruning so that it pays enough time dividend. In many cases, this is well possible, or, abiding by Steven Skiena's famous words: "Clever pruning can make short work of surprisingly hard problems" \cite{skiena1998algorithm}. 

For the Hamiltonian cycle problem, search pruning comes in two forms. The first, \textbf{edge pruning}, involves removing edges from the graph, which can be done either in the preprocessing stage or during a recursive step. By removing edges that \textit{cannot possibly be in any Hamiltonian cycles}, the nodes in the recursive search tree get a lower degree, leading to fewer recursions. Interestingly enough, the prunative removal of edges is always related to the existence of \textit{required edges} that are adjacent to a vertex with degree two and therefore \textit{must} be in any Hamiltonian cycle that might exist in the graph. It can be slightly confusing to keep search pruning and edge pruning apart, but it can be remembered like this: edge pruning means cutting away edges, search pruning involves speeding up the search process. Edge pruning therefore, partially instantiates search pruning.

From literature, we find exactly three edge pruning methods which are all implemented in this study. The first, \textbf{neighbour pruning}, finds a vertex with two required edges, and removes all others. Second, the \textbf{path pruning} method looks for paths of required edges, which might eventually become a Hamiltonian cycle, and removes edges that would prematurely close it. Third, \textbf{solution pruning} removes all edges from the second-last vertex of the partial Hamiltonian path so far. Since path pruning and neighbour pruning might reciprocally facilitate each other's operation, both are repeated until no further pruning occurs. Finally, edges that are removed during preprocessing before recursing are definitely gone, but edges pruned in the recursive step \textit{have} to be put back when the algorithm backtracks.

\subsection{Search Pruning: Non-Hamiltonicity Checks}
  
\noindent The second category of search pruning does not involve the removal of edges, but checking whether a Hamiltonian cycle is achievable at all. There are four such \textit{non-Hamiltonicity checks} and it should be noted that indeed all of them are negative: they can only decide graphs to be either `surely non-Hamiltonian' or `undecided'. Although checks that give `surely Hamiltonian' or `undecided' are certainly imaginable, they are not described in the literature regarding the six general backtrack algorithms used in this study.

The first is a plain and simple \textbf{degree check} which verifies whether any vertex in the graph has edge degree one or zero. If this is the case, the graph cannot be Hamiltonian, and the algorithm needs to backtrack. The second graph configuration that cannot contain a Hamilton Cycle, is a graph where required edges form a closed loop smaller than $v$. This graph configuration can be checked by iterating over all required edges and thus required $O(v)$ computational effort. We call this check the \textbf{premature closure} check. The third check, the \textbf{disconnectedness check} determines whether a graph is broken up into two or more pieces. The subroutine involves picking a vertex and adding it to an empty list of found vertices. From that list, it picks the next item and adds all its adjacent vertices to end of the list, unless they are already added. When the subroutine reaches the end of the list, it counts the number of items in the list. Only if it is equal to $v$, the graph is connected. Note that in some cases, graphs filtered out by this subroutine can also be filtered out by the degree check; it could be considered a hint that the order in which search pruning subroutines are applied also matters for the computation time. The fourth and final check is the  \textbf{one-connectedness check}. A graph is one-connected if it contains a vertex that if removed, breaks up the graph into two or more disconnected parts. Such a vertex is commonly dubbed an \textit{articulation point}, and Tarjan's algorithm finds all articulation points in $O(|v|+|e|)$ time \cite{tarjan1972depth}. To construct a Hamilton Cycle,  there need to be at least two edge-disjoint paths between any two non-adjacent vertices. Therefore, there cannot be a Hamilton cycle in a 1-connected graph. Rubin's is the only algorithm that deploys this technique, we do not know how, but we assumed it could have been done with Tarjan's algorithm, which was published two years earlier. In any case, that is how \textit{we} implemented the check, as literature gave us no definitive answer.

\section{Three Advanced Algorithms}
  
\noindent \textbf{Martello's algorithm \textit{'595'}} is a rare example of an algorithm from the 80's that actually came with written source code in FORTRAN when published. The practice of supplying 'open source' code along scientific experiments is a practice that is much valued today, but was definitely ahead of time in 1983. Martello's algorithm was designed with the purpose of finding \textit{all} Hamiltonian cycles in \textit{directed} graphs. We made the smallest possible changes, adapting the algorithm to undirected graphs, and made it halt after the first solution. Martello’s algorithm has both solution pruning and path pruning, the latter of which is repeatedly applied in each recursion. Furthermore, it uses the $low$ degree preference in its branching heuristic, preferring sparselier connected vertices over denslier connected vertices. 

Like Silvano Martello, Frank Rubin was an influential researcher on early algorithms for NP-complete problems and like Silvano Martello, he designed an algorithm for the \textit{directed} Hamiltonian cycle problem, which we adapted to undirected graphs. \textbf{Rubin's algorithm} was quite sophisticated for the time (1974) and in retrospect would have outperformed most other algorithms. It deploys solution,  path and neighbour pruning exhaustively in each recursion, and additionally performs all four checks for non-Hamiltonicity.

The premature closure check in this algorithm is worth giving some thought, as it might be redundant in combination with the check for disconnectness. We think that if a set of required edges forms a closed cycle smaller than $v$, the graph is automatically disconnected. Furthermore, this algorithm is the only one that performs a check for one-connectedness. In the original paper it is not specified how this check is done, but as stated earlier, it is well possible that this was Tarjan's algorithm and in any case, we implemented it as such. Finally, Rubin does not fully specify his disconnectedness check, but does say it runs in quadratic time. So does ours, and chances are they are nearly identical.

These combined features make it a very advanced algorithm for the time. In fact, no other algorithm in our ensemble deploys so many different subroutines. The only thing it does not employ however, is a branching heuristic. It assumedly just picks the next vertex from the data structure. We think that if Frank Rubin would have just implemented the low-degree branching heuristic too, it would have been the most efficient backtracking algorithm to date. And given that it was contrived nearly half a century ago, these minute details could have profoundly changed the course of history for this problem.

The final algorithm included in this research is the algorithm by Vandegriend and Culberson, abbreviated to \textbf{Vacul's algorithm} \cite{vandegriend1998gn}. It was designed for the undirected Hamiltonian cycle problem and uses solution pruning, path pruning, and neighbor pruning, the last two of which are run exhaustively each recursion. It also uses the degree check, the disconnectedness check and the one-connectedness check, 24 after Frank Rubin introduced them in his algorithm. Since Rubin's study is not referenced by VaCul, it is possible they invented these routines themselves. We can only guess the reasons, but the lack of widely accessible internet papers at the time is at least one likely culprit.

It is sad to see that history missed so many beats, but again, these were the days from before internet and the synchronisation of information; now is the time to make up for it. Important to note is that Vacul's algorithm has an additional feature, a \textit{random restart}, which we left out. Even though stochastically speaking, random restarts on long backtrack runs can save computation time, it makes the algorithm a lot harder to compare to the other algorithms. Vacul's paper comes with a quantitative test on a set of random graphs of variable average degree.

\section{Experiment and results}
\noindent Analogous to previous studies by Cetal, Vacul and Van Horn et al., three sets of randomly generated undirected graphs were created: one with $v=16$ vertices, one with $v=24$ vertices and one with $v=32$ vertices (we will only show and discuss $v=32$, results are comparable for other values of $v$). We created 20 graphs for every number of edges $e = \{1, 2, 3, ... , \frac{1}{2}v(v-1)\}$, resulting in 2400 graphs to solve for $v=16$. For $v=24$, the procedure resulted in 5520 graphs, and 9920 random graphs were generated for the 32-vertex set. This amounts to a subtotal of 17,840 graphs, each of which has been solved by all six algorithms \textit{twice} - once for recursions and once for system time. This means the whole investigation comprises 214,080 runs and therefore, to keep things a little insightful, we will show and discuss results of the 9920-piece ensemble for $v=32$ only. It should be clearly understood though, that the results obtained from the different algorithms in both the time subexperiment and the recursion subexperiment, came from the \textit{same} 17,840 source graphs, facilitating a direct comparison.

\begin{figure*}[b]
    \centering
    \includegraphics[scale=0.14]{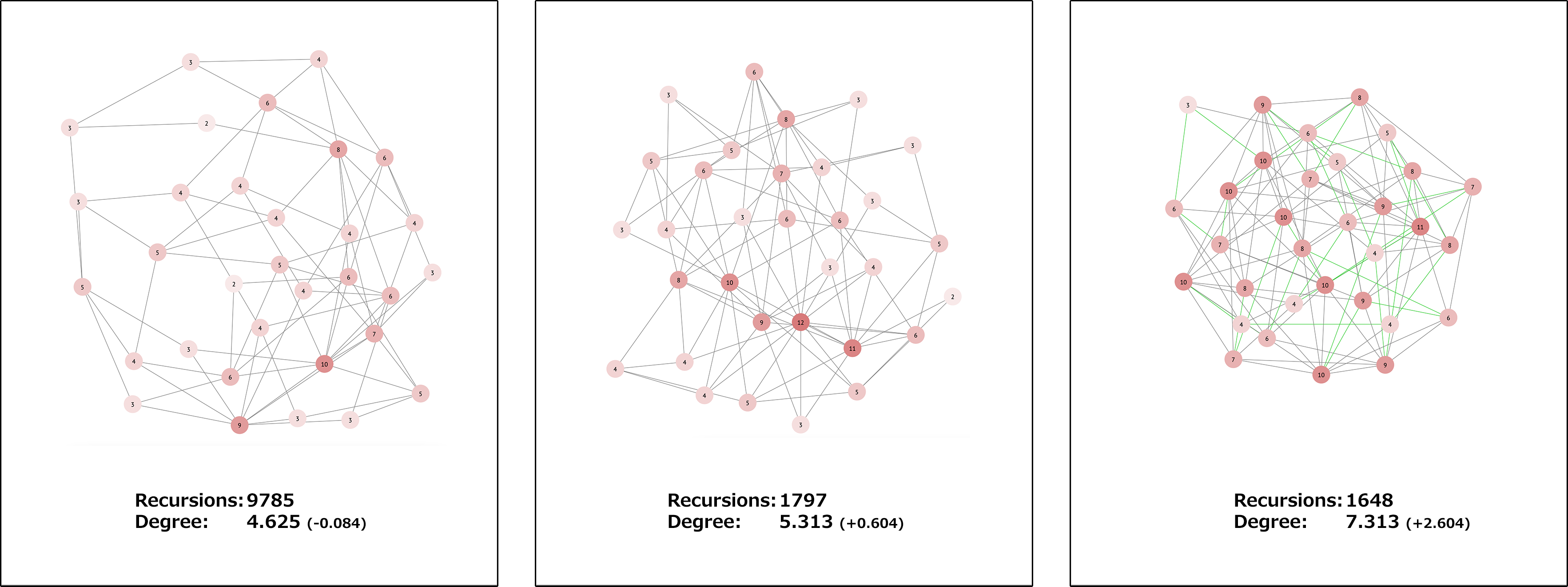}
    \caption{The three hardest graphs for Rubin's algorithm, the best performing algorithm in this study, support the generalized conclusion that hard instances reside near the Koml\'os-Szemer\'edi bound of $ln(32)+ln(ln(32)) \approx 4.71$ (the `distance' in edge degree is in brackets). The suspicion lingers however, that this is only true for \textit{randomly generated} instances ensembles, and targeted results find harder, denser graphs.}
    \label{fig:threehardestgraphs}
\end{figure*}

We first solve (i.e.: decide) all graphs for Hamiltonicity for all six algorithms implemented on a general code base which is publicly accessible \cite{sourcecodejoeri}. Programming the generalized algorithm required some interpretation, as none of the algorithms was published with directly usable source code, but incorporates the three options for branching heuristic, the three edge pruning routines and the four non-Hamiltonicity checks, all in subroutines that can be toggled individually. As such, there is no difference between the branching heuristic in Depth-first and Rubin's, or the the degree-check in Martello's and Vacul's, even if historically speaking, they might not have been identical. The only clearly different feature is the random restart option, which is removed from Vacul's algorithm. There is a lot to say about this procedure, mostly that it \textit{can} be very useful for decision problem instances that \textit{do} have a solution. A somewhat smaller adaptation might be the disconnectedness check and the one-connectedness check, both from Rubin's algorithm. We suspect our quadratic implementation might be conceptually equal to Rubin's, which is also reported as a quadratic order, but eludes further specification. Finally, the sorting procedure for the branching heuristic can be done in various ways whose theoretical and practical runtimes may differ, though not to extend of becoming superpolynomial. We suspect that this routine too, is equal to all in literature, but we can not be sure. Summarizing, all algorithmic components are uniform, deterministic and adapted to undirected graphs. 

We ran two experiments: one recording the number of recursions, and the other recording system time. Traditionally speaking, number of recursions is the way to go. Measuring recursions is immune to the choice of programming language, compiler used, processor speed, parallelization or incoming resource expenditures such as downloads or updates that might interfere with runtimes. Second, the number of recursions computationally speaking largely outweighs pruning and check procedures, which are usually of lower order complexities - typically polynomial versus factorial. Finally, and following previous two reasons, the order complexity of the algorithm also rest with the number of recursions. As, in a broader view, it is the order complexity of algorithms for NP-complete problems that hinge the $P\stackrel{?}{=}NP$ problem, any results which might influence the view on this problem must also be expressed in recursions. However, as we wanted to extend beyond the reach of theoretical computer science and simply get an estimate of time it takes \textit{in the real world} to mobilize the alleged `optimization procedures', and immediately question the `real' impact. A cutoff for the maximum number of a run was $10^5$ recursions, and $10^8$ nanoseconds\footnote{We could have chosen microseconds, milliseconds or even plain seconds here, but this unit holds the best balance between explanation and visualization such as seen in Fig. \ref{fig:zesluik_time}.}

In the recursion experiment, the cutoff was often reached for the three basic algorithms (Fig. \ref{fig:zesluik_recursions}), leaving 20\% of the instances unsolved for depth-first, 8\% for Van Horn's algorithm and a toe-curling 26\% for Cetal's algorithm (for exact values, see Table \ref{table:results_recursions}). Of the more advanced algorithms, Martello's was unable to solve 0.1\% of the instances, but Rubin's and Vacul's sucessfully solved them all. The average number of recursions needed for the solved instances further confirms the hierarchy in the three basic algorithms, and the success of the search pruning efforts. For the advanced algorithms, the ranking slightly changes, with Martello's having the second lowest computational effort within the solved graphs. This changes nothing however for the hierarchy of algorithmic performance when measured in recursions: \textbf{1.Rubin's}, \textbf{2.Vacul's}, \textbf{3.Martello's}, \textbf{4.Van Horn's}, \textbf{5.Depth-first}, \textbf{6.Cetal's}.

For the time experiment, the cutoff value was again reached often for the three basic algorithms (Fig. \ref{fig:zesluik_time}). These results were largely proportional to the recursions-experiment, leaving 19\% of the instances unsolved for depth-first, 8\% for Van Horn's algorithm and a painful 26\% for Cetal's algorithm (for exact values, see Table \ref{table:results_time}). The picture changes slightly for the more advanced algorithms however, where Martello's, Rubin's and Vacul's algorithms left 14, 7 and 4 graphs unsolved, still well below 0.2\% for all three algorithms. Still, it should be noted that \textit{relatively} speaking, this is a huge increase in failure (75\% more failures for Martello's alone). It also swaps the top two contestors in the hierarchy of algorithmic performance relative to the recursion-experiment. The best performing algorithms when measured in system time are: \textbf{1.Vacul's}, \textbf{2.Rubin's}, \textbf{3.Martello's}, \textbf{4.Van Horn's}, \textbf{5.Depth-first}, \textbf{6.Cetal's}.

\section{Conclusion and Discussion}

\noindent It is a fresh awakening that even for graphs of $v=32$, which is relatively small, and the theoretical computational effort required relatively futile, the optimization procedures still make such a large difference. Generally speaking, these procedures appear to pay off in the number of recursions required to decide a graph, especially further away from the Koml\'os-Szemer\'edi bound. When measured in system time however, a different view emerges. The average time to decide a graph hugely increases, especially in denser regions away from the Koml\'os-Szemer\'edi bound.

The big question is how these effects scale up, and whether the increase in search pruning effort really pays off on a larger scale, and whether that counts for recursions, for system time, or for both. Furthermore, the distribution of edge pruning techniques, branching heuristics and non-Hamiltonicity checks among the algorithms is largely haphazard. This is because for now, we tried to stick as closely as possible to historical conventions handed down  to us from literature, but a more structured approach might be desirable. Furthermore, it is quite surprising to see the Rubin's algorithm, which is the second oldest algorithm, overaging Van Horn's by more than four decades, contained the most sophisticated procedures, but also performed best. On the other side, it is a little desillusioning to find Cetal's algorithm, so oftenly cited, finishing last in this test. Finally, some thought must be given to whether other edge pruning techniques, branching heuristics or checks could be devised, and in what order such procedures should be applied. Last but not least, it \textit{might} still pay off to dynamically rearrange the data structure in some way as to make the degree preference in branching dynamically applicable. It should be noted though that this might be programmistically tough, as backtracking would the require un-sorting the dynamically rearranged data structure.

For this study, we made 20 graphs for every possible edge density. Although this procedure is rigourously systematic, it also allows for some skewness in the results: in the end, there is structurally speaking, just one possible graph with one edge. On the other end of the density spectrum, we have 20 `random' graphs with with 496 edges, all being fully connected, and thereby isomorphic. As the number of different graphs for $v$ vertices and $e$ edges is equal to ${\frac{1}{2} v (v-1)}\choose{e}$, the maximum number of different graphs can be found at 248 edges for 32 vertices, which is exactly halfway the Figures \ref{fig:zesluik_recursions} and \ref{fig:zesluik_time}. So if every existible graph of for $v=32$ was equally likely, most of the graphs would a) be Hamiltonian and b) be easy, at least for the best three of our algorithms. For larger graphs, this effect grows stronger as the Koml\'os-Szemer\'edi bound grows (double) logarithmically in $v$, whereas the peak existence grows (half) quadratically. A really weird but inescapable conclusion therefore is, that for larger values of $v$, nearly all instances are Hamiltonian, and many of those might be easy. This does not hold for our results, or that of Cetal, Vacul and Van Horn et al., who all `columnized' their graphs into different degree categories. But combined, the combinatorial assessment and the columnized results show that the prediction of instance hardness for the Hamiltonian cycle problem critically relies on the a priori availability of information of the graph(s) to be solved, even if one future algorithm turns out to be superior.

And this brings us to the last point, tying the discussion back to the preamble. It has been shown that for at least one algorithm, Vacul's in this case, the hardest instances are non-Hamiltonian, very far away from the Koml\'os-Szemer\'edi bound, in a very dense region of the combinatorial space \cite{Sleegers2020Looking}\cite{sleegers2020plant} (also see \cite{vandenbergAdriaans}). This observation apparently contradicts nearly everything that was written in the previous paragraph, but can be explained from the randomness in large ensembles of problem instances such as used in this study. The found extremely hard instances were \textit{structured}, and abiding by the teachings of A. N. Kolmogorov, structured objects in large randomized ensembles are rare \cite{li2008introduction}. So finding these instances paradoxically means knowing where to look, and one way to do that is to use a parameter-unsensitive evolutionary algorithm such as Plant Propagation \cite{salhi2011nature}\cite{Dejonge2020sensitivity}\cite{de2020plant}\cite{paauw2019paintings}\cite{vrielink2019fireworks}\cite{vrielink2021a}\cite{vrielink2021b}. For very large graphs, one could better resort to a single-individual search heuristic, such as HillClimbing or simulated annealing \cite{geleijn2019plant}\cite{Dahmani2020}\cite{kirkpatrick1983optimization}. On the other end of the scale, the use of large randomized ensembles for algorithmic performance raises some questions. Although proper benchmarking is becoming a hot topic \cite{bartz2020benchmarking}, the issue of randomness herein is still seldomly discussed and needs more attention.

\section{Future Work}

\noindent After the results of this study, the road ahead took a few sudden turns. What needs to happen next, is to see if one further generalized backtracking algorithm can convincingly dominate all others. Such a study should include combinations of heuristics and search pruning that are not tested as yet, and it is our belief that such an algorithm exists. After that, its hardest instances should be found, possibly by a global evolutionary search algorithm. Considering earlier results, it could resemble the graph found by Sleegers \& Van den Berg in 2020, as its structure suggests it might be hard for backtracking algorithms in general. Only after that, hardness hierarchies, and possible  implications for computability, and the $P\stackrel{?}{=}NP$ problem itself might be found, but progress in this direction will depend on future developments.

\section{Acknowledgements}

Thanks to Gijs van Horn for maintaining the nice interactive Hamiltonian cycle problem up to date (https://hamiltoncycle.gijsvanhorn.nl/). 







%
%
%

\let\itshape\upshape
\bibliographystyle{IEEEtran}

\end{document}